\documentclass[preprint,journal]{vgtc}       

\ifpdf
  \pdfoutput=1\relax                   
  \usepackage{graphicx}                
  \DeclareGraphicsExtensions{.pdf,.png,.jpg,.jpeg} 
\else
  \usepackage{graphicx}                
  \DeclareGraphicsExtensions{.eps}     
\fi%

\graphicspath{{figures/}{pictures/}{images/}{./}} 

\usepackage{microtype}                 
\PassOptionsToPackage{warn}{textcomp}  
\usepackage{textcomp}                  
\usepackage{mathptmx}                  
\usepackage{times}                     
\usepackage{cite}                      
\usepackage{tabu}                      
\usepackage{xspace}
\usepackage{comment}
\usepackage{enumitem}

\usepackage{ragged2e}

\usepackage{tabularx}

\usepackage{csquotes}
\usepackage{enumitem}

\usepackage{booktabs}
\usepackage{multirow}
\usepackage{graphicx}
\usepackage{soul}

\ieeedoi{10.1109/TVCG.2023.3326904}

\onlineid{1239}

\vgtccategory{Research}
\vgtcpapertype{theoretical \& empirical}

\usepackage{tabularx}
\usepackage{tabulary}

\usepackage{csquotes}

\usepackage{color}

\usepackage{makecell}

\usepackage{lipsum}

\usepackage[normalem]{ulem}
\definecolor{mred}{rgb}{.80,.12,.30}
\definecolor{MRED}{rgb}{.80,.12,.30}
\definecolor{grey}{rgb}{0.5,0.5,0.5}
\definecolor{purple}{rgb}{.75,0,.85}
\definecolor{pistachio}{rgb}{0.58, 0.77, 0.45}

\newif\ifnotes
\notestrue

\let\origcite\cite
\renewcommand{\cite}[1]{\ifnotes\mbox{\origcite{#1}}\else \origcite{#1}\fi}

\title{Knowledge Graphs in Practice: Characterizing their\\Users, Challenges, and Visualization Opportunities}

\author{%
  \authororcid{Harry Li}{0000-0002-2288-6039},
  \authororcid{Gabriel Appleby}{0000-0003-2436-2121},  
  \authororcid{Camelia Daniela Brumar}{0000-0002-7924-634X}, 
  \authororcid{Remco Chang}{0000-0002-6484-6430}, 
  \authororcid{Ashley Suh}{0000-0001-6513-8447}
}

\authorfooter{

    \item Harry Li and Ashley Suh are with MIT Lincoln Laboratory and Tufts University. E-mail: \{harry.li, ashley.suh\}@ll.mit.edu.

  \item Gabriel Appleby, Camelia Daniela Brumar, and Remco Chang are with Tufts University. E-mail: \{gabriel.appleby, camelia\_daniela.brumar, remco.chang\}@tufts.edu.

}

\shortauthortitle{Li et al: Knowledge Graph Interview Study}


\begin{document}

\abstract{
This study presents insights from interviews with nineteen Knowledge Graph (KG) practitioners who work in both enterprise and academic settings on a wide variety of use cases. 
Through this study, we identify critical challenges experienced by KG practitioners when creating, exploring, and analyzing KGs that could be alleviated through visualization design.
Our findings reveal three major personas among KG practitioners -- KG Builders, Analysts, and Consumers -- each of whom have their own distinct expertise and needs. We discover that KG Builders would benefit from schema enforcers, while KG Analysts need customizable query builders that provide interim query results. 
For KG Consumers, we identify a lack of efficacy for node-link diagrams, and the need for tailored domain-specific visualizations to promote KG adoption and comprehension.
Lastly, we find that implementing KGs effectively in practice requires both technical and social solutions that are not addressed with current tools, technologies, and collaborative workflows. 
From the analysis of our interviews, we distill several visualization research directions to improve KG usability, including knowledge cards that balance digestibility and discoverability, timeline views to track temporal changes, interfaces that support organic discovery, and semantic explanations for AI and machine learning predictions.
}



\maketitle


\section{Introduction}
\label{sec:introduction}

Knowledge graphs have emerged as a popular approach to represent and manage complex data from a variety of domains\cite{abu2021domain}. 
%
Due to their ability to encode semantically rich information in the form of entities and the relationships between them, knowledge graphs are now an industry standard for data unification\cite{xiao2019virtual}, question-answering\cite{huang2019knowledge}, recommendation systems\cite{guo2020survey}, explainable AI\cite{lecue2020role}, and many other practical applications\cite{hogan2021knowledge}.
However, despite the growing popularity of knowledge graphs (KGs), there is a limited understanding of the \textit{types} of KG users, the challenges they face, and the limitations of current tools and visualization designs for KGs used in practice.

To address this gap, we conducted an interview study with 19 KG practitioners across eight different organizations. The participants of our study come from a broad background of both industry and academic experiences, representing a diverse set of domains -- including biology, finance, health, drug development, software, cybersecurity, information science, and materials science. From the analysis of our interviews, we provide a characterization of the common personas of KG users, their expertise, tool usage, the obstacles they face when using KGs, and their unmet visualization needs. 
We then propose new directions for visualization research that leverages the semantic richness of KGs to both improve upon existing designs and address common challenges.


We identify three personas for KG users: \textit{Builders} who construct and maintain KGs, \textit{Analysts} who explore and extract insights from KGs, and \textit{Consumers} who use insights from KGs for downstream tasks and decision-making. Across these three personas, we find that their expertise, tasks, and needs can vary drastically. 
Common challenges experienced by these personas include: difficulty querying KGs, poor data quality, evolving and mismanaged data provenance, schema inconsistencies, and a lack of organizational KG standardizations.

Several participants stressed sociotechnical challenges in understanding the desired outcomes of a knowledge graph.
When this happens, Builders can overcomplicate the construction of the KG 
(e.g., with too many features), rendering it unsustainable. Analysts can struggle to acquire and deliver relevant insights to stakeholders, while Consumers may ultimately perceive no utility in the KG's use for their downstream tasks. Consequently, organizations fail to adopt the KG in practice.

In addition to sociotechnical challenges, issues were raised related to current visualization methods. These challenges include: scalability, limited support for organic discovery, unaddressed domain-specific needs, and overly complex visualizations for end users. We find that node-link diagrams, although frequently used as a visual medium, are often ineffective for both generating and delivering insights from large KGs. 
For instance, one practitioner told us their software team's end users ultimately preferred simple table-based KG representations over custom-built interactive graph interfaces.
Overall, we identify a need for visualization solutions that are targeted towards \textit{each} KG persona, particularly those employing higher levels of abstraction to facilitate communication with diverse audiences.

When asked about what types of visual interfaces could be beneficial for exploring KGs, we found that participants widely praised Wikipedia's ability to support insight generation through on-the-fly data and entity hopping. 
In particular, ``Wikipedia-style'' interfaces are desirable for users when
engaging in \textit{open-ended KG exploration}, where there is no specific analysis target. 
Conversely, search engines like Google can help users when they want to pose direct and precise questions to the KG, representing \textit{goal-oriented KG exploration}. 
As a whole, participants emphasized that there is currently no universally accepted solution for either type of KG-based exploratory analysis.


The remainder of our paper is structured as follows: Section~\ref{sec:related} provides a background on knowledge graphs,
as well as previous work on visualization practices for KGs. Section~\ref{sec:methodology} outlines our protocol and analysis methodology for our interviews with KG practitioners. In Sections~\ref{sec:findings-kg-users},~\ref{sec:findings-kg-challenges}, and~\ref{sec:findings-kg-opportunities}, we respectively characterize the users, challenges, and visualization opportunities of KGs based on our findings. Finally, in Section~\ref{sec:discussion} we discuss the limitations and future work for this study.

To summarize, the major contributions of this paper are:

\begin{itemize}[topsep=1pt, partopsep=0pt,itemsep=2pt,parsep=0pt]
    \item A thematic analysis of interviews conducted with 19 KG practitioners across eight different organizations who regularly create, explore, analyze, and deliver insights from KGs. 
    \item A characterization of KG users, their common organizational roles, areas of expertise, tool usage, and visualization needs. 
    \item Directions for future KG visualization research, along with design sketches iterated on with domain experts, aimed at alleviating KG-related challenges identified from our interview study.
\end{itemize}

\begin{figure} 
    \centering
     \includegraphics[clip, trim=0 3.8cm 6.7cm 0, width=.45\textwidth]{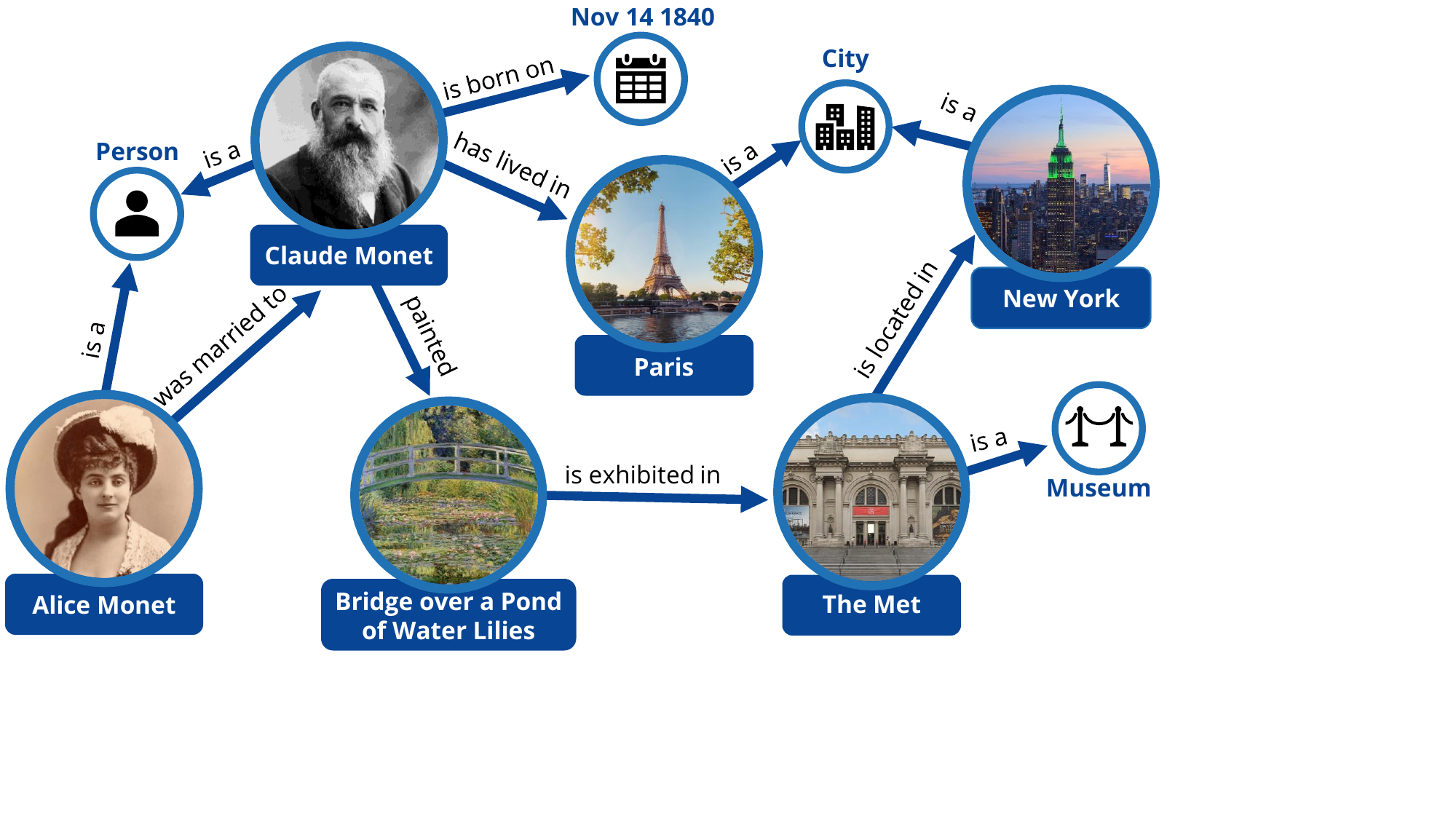} 
    \caption{%
    An illustrative example of a knowledge graph (KG). In a KG, different types of entities (nodes) can have different types of relationships (edges) defined between them. We further discuss KGs in Section~\ref{sec:background}. 
    }
    \label{fig:kg-example}
\end{figure}

\section{Background \& Related Work}
\label{sec:related}
We begin with a brief background on knowledge graphs and introduce terminology used throughout this paper. We then discuss related work in KG system design, as well as KG visualization tools and practices.

\subsection{Knowledge Graphs}
\label{sec:background}

Fundamentally, a knowledge graph is a data model that represents knowledge in the form of nodes (i.e. \textit{entities}), edges (i.e. the \textit{relations} between entities) and properties (i.e. \textit{attributes}) that can be defined for both nodes and edges. A visual illustration of a simple KG can be seen in Figure~\ref{fig:kg-example}. 
The concept of a knowledge graph dates back to the rise of the semantic web\cite{berners2001semantic} in 2001; however, KGs gained notable popularity after the launch of Google's Knowledge Graph\cite{ehrlinger2016towards} in 2012, along with the demand for more sophisticated representations of data.

KGs store their information as triplets (e.g., \{\textit{head, relation, tail}\}, or \{\textit{subject, predicate, object}\}), providing a robust (and often hierarchical) structure to reason about data. 
The entire ``blueprint'' that defines the KG's structure (e.g., its nodes, edges, properties) is its \textit{schema}. A schema specifies the different node types, edge types, properties, and constraints in the KG. By doing so, the schema ensures consistency, thereby facilitating data integration and providing a shared understanding of the data model for effective querying and analysis.

There are several well-known knowledge graph databases commonly used today (e.g., Stardog\cite{union2020stardog}, Neo4j\cite{miller2013graph}) with respective graph query languages (e.g., SPARQL\cite{ell2014sparql}, CYPHER\cite{francis2018cypher}). 
By leveraging the underlying structure of the graph, queries on knowledge graphs can retrieve specific relationships, navigate through different nodes, and uncover meaningful patterns in the data. Graph analytic algorithms can also be applied to generate new features, as well as to learn and make predictions about data\cite{Bonatti:2019:Knowledge}. This enables knowledge graphs to support classification and regression analysis on both its nodes and edges.

The semantic nature of KGs makes them well-suited for enhancing language tasks, particularly when combined with large language models (LLMs)\cite{alkhamissi2022review, petroni2019language, brown2020language}.
KGs are also used to manage diverse data sources, including data lakes, data warehouses, and knowledge bases (e.g., WikiData).
For a complete background on KGs, including their common applications, we point to Hogan et al.'s survey paper\cite{hogan2021knowledge}.

\subsection{Studies in Knowledge Graphs}

Understanding the benefits, shortcomings, and future research directions for knowledge graphs was outlined in a 2019 Dagstuhl seminar report\cite{Bonatti:2019:Knowledge}. In the same year, another Dagstuhl report provided new directions in visual analytics research for multilayer networks\cite{kivela2019visual}, which have similar properties to KGs. A 2022 Dagstuhl report on the intersection of graph databases and network visualization\cite{klein2022bringing} is likely the most similar in motivation to our work here. 

However, to the best of our knowledge, our interview study is the first attempt to characterize the practitioners of KGs, their challenges, their tool usage, and their visualization needs. Similar qualitative work has been done within the visualization research community to understand the obstacles faced by ML experts\cite{passi2018trust, hong2020human} (e.g., \textit{data cascades}\cite{sambasivan2021everyone}), client-facing data scientists\cite{mosca2019defining}, exploratory data analysts\cite{kandel2012enterprise, alspaugh2018futzing, wongsuphasawat2019goals}, and ML stakeholders\cite{suresh2021beyond}. In addition to improving collaborations, these studies are similarly conducted to discover new opportunities for visualization to alleviate challenges faced by practitioners. 

A recent position paper by Lissandrini et al.\cite{lissandrini2022knowledge} called for better KG exploration tools. The authors highlight important tasks and use cases for KG systems, particularly for KG creators and maintainers. From our interviews, we identify similar challenges and needs for users (e.g., maintaining KG schemas, scalability, and the demand for interim query results); however, we also identify unmet visualization needs for KG builders, analysts, and consumers. While the authors provide an excellent foundation for database system designs, our work differs in that we interview KG practitioners directly to understand their needs across various technologies and visualization tools.


\subsection{Visualization Solutions for Knowledge Graphs}
\label{sec:visualization-solutions-for-KGs}
There are an increasing number of systems that aim to visualize and explore the data in knowledge graphs.
Latif et al. contributed \textit{VisKonnect}\cite{latif2021visually}, a multi-coordinate visualization system for EventKG\cite{gottschalk2018eventkg} that analyzes the connections of historical figures based on the events they participated in. 
VisKonnect includes an NLP-based panel that lets users ask templated questions to the KG 
with the GPT-3 language model\cite{brown2020language}.
%
%
Ahmad et al.\cite{ahmad2021towards} contributed a visualization that maps data from a KG for patients with inflammatory bowel disease
to compare their history, progressions, and administered treatments. 
Husain et al. contributed a multi-scale visual analytics approach for exploring biomedical knowledge graphs\cite{husain2021multi}. 
Their approach includes three types of views, a `global' (high-level) view, a local `drilled down' view, and a text-evidence document view.
Partl et al.\cite{2016_eurovis_pathfinder} contributed a scalable path finding approach with multiple views that queries and ranks candidate paths by topological features, as well as node and edge properties.
While more visualization tools are being designed for and around KGs, they are not without their own set of limitations\cite{lissandrini2022knowledge}. KGs are typically very large in size, may have repeat entities or attributes with similar names (i.e.\ entity ambiguation\cite{Gao2015DataToneMA}), and may contain obsolete or out-of-date data\cite{hogan2021knowledge}. 
For this paper, we target the identification of similar issues preventing wider visualization adoption for KGs, and potential opportunities to alleviate those challenges. 



\subsection{Knowledge Graph Solutions for Visual Analysis}
In addition to visualization solutions targeted for knowledge graphs, the research community has looked into how we can leverage KGs to build better visualizations and interfaces.
Dating back to 2008, Chan et al. presented Vispedia\cite{chan2008vispedia}, an interactive visual exploratory tool that allows users to integrate and visualize data tables from DBpedia\cite{auer2007dbpedia}.
%
KG4VIS\cite{li2021kg4vis}
recommends
visualizations using a knowledge graph constructed from a large corpus of data-visualization pairs. KG4VIS can also generate
rules from a KG embedding to ``explain'' why the model recommends certain visualizations given the user's data.
Cashman et al.'s CAVA system\cite{Cashman:2020:CAVA} utilizes KGs to help users interactively perform data augmentation on their existing datasets. Specifically, users are able to automatically join new attributes from a KG to improve performance on analysis tasks, e.g., improving a model's predictive power.

While many of the tools discussed in this section are robust to their particular data domain and use case, it is currently unclear whether current solutions and visualization techniques can cover broader KG practitioner needs. In the following section, we describe an interview study to investigate these questions. In Section~\ref{sec:node-link-diagrams}, we distill the potential benefits and tradeoffs of current KG visualization designs.


\begin{table*}[ht!]
    \centering
    \renewcommand{\arraystretch}{1.3}
    \centering
    \resizebox{.95\linewidth}{!}{%
    \definecolor{palesilver}{rgb}{0.9, 0.9, 0.9}

\renewcommand\theadalign{bt}
\renewcommand\theadfont{\bfseries}
\renewcommand\theadgape{\Gape[4pt]}
\renewcommand\cellgape{\Gape[4pt]}

\sffamily
\resizebox{\textwidth}{!}{
\begin{tabular}[t]{lllllclllll}
\toprule

\textbf{PID} & 
\textbf{Education} & 
\textbf{Job Title} & 
\textbf{Company Domain} & 
\textbf{KG Persona(s)} & 
\thead{Years of\\Experience} & 
\thead{Familiarity\\with KGs} &
\thead{Familiarity\\Creating KGs} &
\thead{Familiarity\\Analyzing KGs} &
\thead{Familiarity\\Querying KGs} &
\thead{Familiarity\\Visualizing KGs} \\
\midrule

01  & MS & Research Scientist & FFRDC & Builder, Analyst & 4 & 4 \textcolor{grey}{\textcolor{grey}{\textcolor{grey}{(Moderate)}}} & 4 \textcolor{grey}{\textcolor{grey}{(Moderate)}} & 4 \textcolor{grey}{\textcolor{grey}{(Moderate)}} & 3 \textcolor{grey}{(Some)} & 3 \textcolor{grey}{(Some)} \\
02  & MS & Research Scientist & FFRDC & Builder, Analyst & 20 & 3 \textcolor{grey}{(Some)} & 3 \textcolor{grey}{(Some)} & 3 \textcolor{grey}{(Some)} & 3 \textcolor{grey}{(Some)} & 4 \textcolor{grey}{\textcolor{grey}{(Moderate)}} \\
03  & PhD & Research Scientist & FFRDC & Analyst & 2 & 3 \textcolor{grey}{(Some)} & 3 \textcolor{grey}{(Some)} & 3 \textcolor{grey}{(Some)} & 3 \textcolor{grey}{(Some)} & 3 \textcolor{grey}{(Some)} \\
04  & MS & Research Scientist & FFRDC & Analyst & 5 & 4 \textcolor{grey}{\textcolor{grey}{(Moderate)}} & 3 \textcolor{grey}{(Some)} & 4 \textcolor{grey}{\textcolor{grey}{(Moderate)}} & 3 \textcolor{grey}{(Some)} & 3 \textcolor{grey}{(Some)} \\
05  & MS & Research Scientist & FFRDC & Builder, Analyst & 3 & 4 \textcolor{grey}{\textcolor{grey}{(Moderate)}} & 4 \textcolor{grey}{\textcolor{grey}{(Moderate)}} & 4 \textcolor{grey}{\textcolor{grey}{(Moderate)}} & 3 \textcolor{grey}{(Some)} & 3 \textcolor{grey}{(Some)} \\
06  & MS & Research Scientist & FFRDC & Analyst & 5 & 3 \textcolor{grey}{(Some)} & 3 \textcolor{grey}{(Some)} & 3 \textcolor{grey}{(Some)} & 3 \textcolor{grey}{(Some)} & 2 \textcolor{grey}{(Slight)} \\
07  & MS & Research Scientist & FFRDC & Analyst & 5 & 4 \textcolor{grey}{\textcolor{grey}{(Moderate)}} & 4 \textcolor{grey}{\textcolor{grey}{(Moderate)}} & 4 \textcolor{grey}{\textcolor{grey}{(Moderate)}} & 3 \textcolor{grey}{(Some)} & 2 \textcolor{grey}{(Slight)} \\
08  & MS & Software Developer & FFRDC & Builder, Analyst & 3 & 2 \textcolor{grey}{(Slight)} & 2 \textcolor{grey}{(Slight)} & 2 \textcolor{grey}{(Slight)} & 3 \textcolor{grey}{(Some)} & 2 \textcolor{grey}{(Slight)} \\
09  & MS & Research Scientist & FFRDC & Builder, Analyst & 2 & 4 \textcolor{grey}{\textcolor{grey}{(Moderate)}} & 5 \textcolor{grey}{(Extreme)} & 3 \textcolor{grey}{(Some)} & 4 \textcolor{grey}{\textcolor{grey}{(Moderate)}} & 2 \textcolor{grey}{(Slight)} \\
10 & PhD & Research Scientist & FFRDC & Builder, Analyst & 5 & 3 \textcolor{grey}{(Some)} & 2 \textcolor{grey}{(Slight)} & 3 \textcolor{grey}{(Some)} & 4 \textcolor{grey}{\textcolor{grey}{(Moderate)}} & 3 \textcolor{grey}{(Some)} \\
11 & MS & Data Analyst & Enterprise (Finance) & Builder, Analyst & 2 & 4 \textcolor{grey}{\textcolor{grey}{(Moderate)}} & 4 \textcolor{grey}{\textcolor{grey}{(Moderate)}} & 4 \textcolor{grey}{\textcolor{grey}{(Moderate)}} & 3 \textcolor{grey}{(Some)} & 1 \textcolor{grey}{(None)} \\
12 & PhD & Director & Enterprise (Health) & Builder, Analyst & 8 & 4 \textcolor{grey}{\textcolor{grey}{(Moderate)}} & 4 \textcolor{grey}{\textcolor{grey}{(Moderate)}} & 4 \textcolor{grey}{\textcolor{grey}{(Moderate)}} & 4 \textcolor{grey}{\textcolor{grey}{(Moderate)}} & 4 \textcolor{grey}{\textcolor{grey}{(Moderate)}} \\
13 & BS & Industry Analyst & Enterprise (Consulting) & Analyst, Consumer & 2 & 5 \textcolor{grey}{(Extreme)} & 3 \textcolor{grey}{(Some)} & 3 \textcolor{grey}{(Some)} & 3 \textcolor{grey}{(Some)} & 4 \textcolor{grey}{\textcolor{grey}{(Moderate)}} \\
14 & MS & PhD Student & Academia & Builder, Analyst & 5 & 4 \textcolor{grey}{\textcolor{grey}{(Moderate)}} & 4 \textcolor{grey}{\textcolor{grey}{(Moderate)}} & 4 \textcolor{grey}{\textcolor{grey}{(Moderate)}} & 4 \textcolor{grey}{\textcolor{grey}{(Moderate)}} & 4 \textcolor{grey}{\textcolor{grey}{(Moderate)}} \\
15 & PhD & Data Scientist & Enterprise (Health) & Analyst & 4 & 4 \textcolor{grey}{\textcolor{grey}{(Moderate)}} & 2 \textcolor{grey}{(Slight)} & 5 \textcolor{grey}{(Extreme)} & 5 \textcolor{grey}{(Extreme)} & 4 \textcolor{grey}{\textcolor{grey}{(Moderate)}} \\
16 & PhD & Comp. Biologist & Enterprise (Health) & Builder, Analyst & 5 & 4 \textcolor{grey}{\textcolor{grey}{(Moderate)}} & 5 \textcolor{grey}{(Extreme)} & 4 \textcolor{grey}{\textcolor{grey}{(Moderate)}} & 4 \textcolor{grey}{\textcolor{grey}{(Moderate)}} & 4 \textcolor{grey}{\textcolor{grey}{(Moderate)}}\\
17 & PhD & Principal Scientist & Enterprise (Tech) & Builder, Analyst & 15 & 5 \textcolor{grey}{(Extreme)} & 5 \textcolor{grey}{(Extreme)} & 5 \textcolor{grey}{(Extreme)} & 5 \textcolor{grey}{(Extreme)} & 4 \textcolor{grey}{\textcolor{grey}{(Moderate)}}\\
18 & MBA & Digital Lead & Enterprise (Health) & Consumer & 1 & 3 \textcolor{grey}{(Some)} & 2 \textcolor{grey}{(Slight)} & 2 \textcolor{grey}{(Slight)} & 2 \textcolor{grey}{(Slight)} & 3 \textcolor{grey}{(Some)} \\
19 & MS & PhD Student & Academia  & Builder & 2 & 4 \textcolor{grey}{\textcolor{grey}{(Moderate)}}& 4 \textcolor{grey}{\textcolor{grey}{(Moderate)}} & 3 \textcolor{grey}{(Some)} & 3 \textcolor{grey}{(Some)} & 2 \textcolor{grey}{(Slight)}\\



\bottomrule

\end{tabular}}}
    \caption{Participant demographics for our interview study, described in Section~\ref{sec:methodology}. From left to right: the participant's ID; job title; the organization they work in (FFRDC stands for Federally Funded Research and Development Center); their primary persona(s) as KG users (further explained in Section~\ref{sec:kg-roles}); years of experience with KGs; overall familiarity working with KGs, creating or maintaining KGs, exploring or analyzing KGs, querying KGs, and visualizing KGs. Familiarity was self-reported on a Likert Scale from (1) not at all familiar to (5) extremely familiar. }
    \label{tab:participant-demographics}
\end{table*}

\section{Methodology}
\label{sec:methodology}
To better understand the users of KGs, their use cases, challenges, and visualization needs, we conducted an interview study with KG practitioners from both research and enterprise settings. 
All interview and supplemental materials can be accessed at \url{https://github.com/TuftsVALT/KGsInPractice}.

\subsection{Participant Recruitment}
We recruited interview participants via emails to 
the authors' professional contacts. In the recruitment email, we specified that interviews would be conducted with practitioners 
who had experience working with knowledge graphs in some capacity (e.g., creating, maintaining, querying). The demographics for our final 19 interview participants is shown in Table~\ref{tab:participant-demographics}. P1-10 come from varying divisions within the same FFRDC. P12, P15, and P16 come from the same company. The remaining participants are from 6 different organizations.

Participants self-reported their demographics, including their highest education level, job title, company domain, and subjective familiarity with knowledge graphs. Familiarity was selected on a Likert \textit{Level of Familiarity} Scale from 1-5, where 1=Not at all familiar, 2=Slightly familiar, 3=Somewhat familiar, 4=Moderately familiar, and 5=Extremely familiar. The following questions were asked regarding familiarity: 

\begin{enumerate}[topsep=2pt, partopsep=0pt,itemsep=1pt,parsep=2pt]
\item How would you rate your familiarity with KGs \textbf{in general}?
\item How would you rate your familiarity \textbf{creating} or \textbf{modifying} KGs?
\item How would you rate your familiarity \textbf{exploring}, \textbf{analyzing}, and \textbf{gaining insights} from KGs?
\item How would you rate your familiarity \textbf{querying} KGs?
\item How would you rate your familiarity \textbf{visualizing} KGs?
\end{enumerate}


\subsection{Protocol}
Two authors conducted all interviews during a six month period. Most (18/19) interviews were conducted virtually through video conferencing software, with one conducted in person. We recorded and transcribed 15/19 interviews, and took detailed notes for the remaining where recording was not possible due to the interviewee's company policies. 

Each interview lasted roughly one hour. 
Three of our interviews were conducted as focus groups (Group 1: P2, P3, P4; Group 2: P5, P6; Group 3: P9, P10) while the remaining were individual interviews. 
Participants were walked through the same set of PowerPoint slides prepared by the authors. This slide deck contained questions to help elicit the participants' KG experience, roles, projects, challenges, and visualization needs. The slides are provided as supplemental material.

After each interview, the authors met to discuss emergent themes, following a thematic analysis process\cite{braun2006using}. We determined our sample size by reaching thematic saturation, where consistent themes emerged, and no new findings or potential codes were identified. In the end, we concluded our study after interviewing a total of 19 KG practitioners.

%

\subsection{Interviews}
\label{sec:methodology-interviews}
At the start of each interview, participants were given a high-level definition of a knowledge graph and an illustrative example image (Figure~\ref{fig:kg-example}) of one containing historical figures as nodes, relationships between figures as edges, and attributes belonging to both.  
In Section~\ref{sec:discussion-kg-definition} we discuss our participants' own definitions of a knowledge graph. 

Following this briefing, slides were shown asking participants to describe their experience with KGs: how they were created, how they were queried, maintained, and their characteristics (e.g., size, schema, domain). We asked which types of questions participants were trying to answer with the KG, which data domain they worked in, which tools were commonly used, and which (if any) visualization solutions or tools had been helpful in the past.


To summarize, the overarching questions for our interviews were:

\begin{enumerate}[topsep=2pt, partopsep=0pt,itemsep=1pt,parsep=2pt]
\item What is your experience with KGs (what do you use them for, how are they created, what are common challenges faced)?
\item What kinds of questions do you try to answer with your KGs?
\item What tools or techniques do you use, and what is the typical outcome of using them?
\item What visualization tools or methods do you use for KGs? What do you look for in a visualization tool, and what is missing?
\item At what granularity do you want to see or view the KG? 
\end{enumerate}

We also walked participants through several examples of knowledge graph visual analysis tools (specifically,\cite{latif2021visually, gottschalk2018eventkg, neo4jbloom}) to understand what could be potentially helpful or not helpful for different KG use cases. Participants' feedback, in addition to their responses to the above questions, helped to inform the opportunities for future KG visualization research we distill in Section~\ref{sec:findings-kg-opportunities}. A write-up of their full responses to the presented tools is provided in our supplemental.

\subsection{Analysis}
The goal of our analysis was to qualify the users of KGs, their applications, their frequently experienced challenges, and visualization needs that are not satisfied by current technologies.   
We carried out a qualitative coding process to analyze our interview data. Our codebook was developed iteratively between all authors, and followed the protocols in\cite{macqueen1998codebook, decuir2011developing} for good codebook development. After each interview, the authors met to recap the discussion and note the most common uses cases, tools, issues, and needs experienced by the participant(s). Through this process, the strongest themes that emerged from our analysis related to: (1) why KGs are used over other solutions (\textit{use cases}); (2) tools used for and with KGs (\textit{tools}); (3) problems working with KGs (\textit{challenges}); (4) lack of visualization support (\textit{visualization needs}). 

The interview data was coded by the original interviewers. We determined an utterance in the interviews to be a participant's turn in conversation, where a turn was a response to a single question from the slide deck. An individual code could only be assigned once per utterance, but many codes could be assigned in an utterance.  Both authors coded the set of interviews, swapped to agree or disagree with each other's coding, then swapped again to make resolutions. Our entire coding process was thoroughly collaborative and conducted to holistically categorize the interview findings, rather than rigorously compare the frequency of code usage. Therefore, disagreements were discussed and ultimately decided between both coders. 

To remain consistent with our analysis procedure, we present the qualitative findings of our interviews in the following section in terms of our themes, and only use code counts to report the total number of participants experiencing a particular challenge, use case, and so on. 

 \begin{figure*}[ht!] 
    \centering
     \includegraphics[clip, trim=1pt 12cm 7.8cm 1pt, width=.98\textwidth]{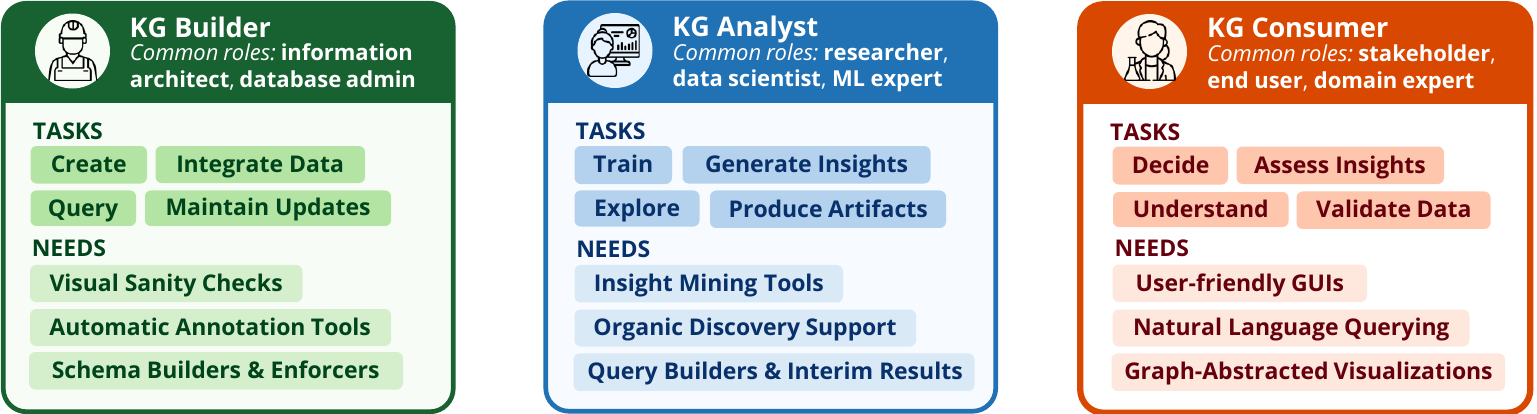} 
    \caption{%
    Three personas we identified for the users of knowledge graphs from our interviews, described in Section~\ref{sec:methodology}. From the left, a user can be a KG Builder (e.g., database administrator), an Analyst (e.g., data scientist), or a Consumer (e.g., stakeholder). All three types of KG users have distinct roles, tasks, needs, and expertise -- however, it is possible a user can belong to more than one persona. For example, a user that creates their own KG of companies (``KG Builder'') to predict which to invest into (``KG Analyst''). We further describe these personas in Section~\ref{sec:kg-roles}.   %
    }
    \label{fig:kg_user_profiles}
\end{figure*}

\section{Knowledge Graph Users \& Current Practices}
\label{sec:findings-kg-users}
First we distill personas for the users of knowledge graphs, their common uses cases, data sources, reasons for using a KG over other data models, as well as their frequently used tools and technologies. 

\subsection{KG Personas}
\label{sec:kg-roles}

From our interviews, we identify three major KG practitioner personas, highlighted in Figure~\ref{fig:kg_user_profiles}. While each persona comes with distinct expertise, responsibilities, and needs, we often found that one person could step into multiple personas depending on their use case or organizational role. We assigned personas to each participant in Table~\ref{tab:participant-demographics} based on the variety of KG tasks they regularly perform.

\begin{itemize}[leftmargin=0pt,topsep=4pt, partopsep=0pt,itemsep=2pt,parsep=2pt]
    
    \item[]{\textbf{KG Builder:} 
    The KG Builder is usually an expert in database systems, data management, or data modeling. Builders are responsible for creating a KG from its source data, deciding which database or representation method to use for storing the KG (discussed more in Section~\ref{sec:tools}), and developing the KG's schema. Builders are typically an expert with one or more graph querying languages, particular those associated with their chosen KG representation. Many (12/19) of our interview participants fit into the Builder persona.}

    \item[]{\textbf{KG Analyst:} 
    The KG Analyst is typically an expert in data science or ML. Analysts are responsible for generating insights from the KG either as artifacts (e.g., reports) for end users, or as input for downstream analysis and AI/ML tasks. KG Analysts have familiarity with extracting information from the KG via its querying language, but may not necessarily be experts. Most of our participants who fit into the Builder persona also fit into the Analyst persona (11/12), as naturally Builders either are themselves the Analysts, or work closely with Analysts to construct the KG to meet the analysis use case. The majority (17/19) of our participants fit into the Analyst persona.}

    \item[]{\textbf{KG Consumer:} 
    The KG Consumer is generally an expert in the data domain, business, overarching use case, or the KG's sociocultural context (i.e. milieu\cite{suresh2021beyond}). While Consumers typically do not interact directly with a KG database or its querying language, they are still a stakeholder or end user of the KG, and 
    know what ``types'' of insights would be valuable to extract. Consumers tend to rely on KG Analysts, query building GUIs, or automated reporting systems to generate those insights. Few (2/19) of our participants fit appropriately into the Consumer persona; we discuss this limitation in Section~\ref{sec:discussion-limitations}.
    }

\end{itemize}

\begin{table}[ht!]
    \centering
    \renewcommand{\arraystretch}{1.2}
    \centering
    \small
    \resizebox{.95\linewidth}{!}{%
    \definecolor{palesilver}{rgb}{0.9, 0.9, 0.9}

\renewcommand\theadalign{bt}
\renewcommand\theadfont{\bfseries}
\renewcommand\theadgape{\Gape[4pt]}
\renewcommand\cellgape{\Gape[4pt]}
\sffamily
\resizebox{\textwidth}{!}{
\begin{tabular}[t]{p{0.25\linewidth}p{0.75\linewidth}}
\toprule

\textbf{Use Case} &
\textbf{Description} \\

\midrule

Data cataloging & Creating a query-able knowledge base for data scientists and developers to quickly find data they need\\
 & Standardizing and de-duplicating terminology and data usage\\
 & Modeling the organization's business logic, e.g. ``the organization has \textit{X} branch, which has \textit{Y} types of employees''\\
  & Modeling facilities, their security systems, lights, fire suppression, etc. to connect to physical floor layouts\\
& Managing global web content, e.g., showing information about a movie based on the website visitor's country\\

 \hline 
 
Path Discovery & Finding new cyber threat pathways in a cyber KG\cite{hemberg2020linking} connecting computer systems and known exploits\\
 & Finding new treatment pathways in a KG connecting diseases and possible treatments (e.g., for drug discovery) \\
 & Finding new materials synthesis pathways in a KG connecting different chemical compositions\\
 & Identifying user workflows in a KG connecting user actions in an enterprise network\\

\hline 

AI/ML & Explaining why an anomaly was predicted from a model using data that is also connected to the KG\\
 & Predicting stock prices for publicly traded companies in a KG connecting companies, industries, and supply chains\\
 & Using NLP to process text-based data sources (social networks)  to detect author profiles and authoring changes \\



\bottomrule
\end{tabular}}}
    \caption{A select set of our participants' KG use cases. KGs are used for data cataloging, pathway discovery, as well as training, understanding, and improving AI/ML models. More details in Section~\ref{sec:users-use-cases}.}
    \label{tab:participant-use-cases}
\end{table}

\subsection{KG Use Cases}
\label{sec:users-use-cases}

Our participants are using KGs for a wide variety of use cases (see Table~\ref{tab:participant-use-cases} for examples, and the supplemental for a full list): 

\begin{itemize}[topsep=1pt, partopsep=0pt,itemsep=1pt,parsep=0pt]
  \item 5/19 use KGs as enterprise data catalogs or data warehouses;
  \item 9/19 use KGs for path discovery;
  \item 15/19 use KGs with AI/ML to improve data quality, to analyze data, and to improve their predictive models. 6/19 specifically use KGs for node classification or regression, i.e. predicting a class label or numeric score for a set of nodes in the graph.
\end{itemize}

In terms of scale, 8/19 of our participants work with KGs containing thousands of nodes, 7/19 work with millions of nodes, and 4/19 work with billions of nodes. The number of edges ranged from millions to billions, with 3 to 10,000 properties on nodes and edges.

\subsection{Benefits \& Affordances of Using KGs}
\label{sec:benefits-of-using-kgs}
We also asked participants why they preferred to use KGs over other data structures, like traditional graphs or relational databases. 

\begin{itemize}[leftmargin=0pt,topsep=4pt, partopsep=0pt,itemsep=2pt,parsep=2pt]
    
    \item[]{\textbf{Schema Flexibility:} KGs were praised as versatile due to their flexible schema structure, particularly when compared to other data structures:

    \begin{displayquote}
        If you think about a relational database, the column number must be the same for all the records, right? But for our KG\ldots
        if you want to add a new type of data to a certain record, you can simply add in an edge. -P19
    \end{displayquote}
    }


    

    \item[]{\textbf{Integration Across Multiple Data Sources 
        \& Domains:} Participants reported integrating both public (13/19) and non-public (10/19) data to generate and contextualize their KGs for end users.
%
        %
        %
        This allows Analysts and Consumers to discover previously unknown relationships:
    
        \begin{displayquote}
        The individual associations are insignificant on their own, it's only when we look at the multiple associations 
        in the context of the graph do we\ldots really find clusters.
        -P18
        \end{displayquote}
    }

    \item[]{\textbf{Semantic Encodings:} A KG's semantic nature goes hand in hand with its robust ability to manage data. One participant told us a KG's ability to encode semantic-based relationships makes its usage worth the additional complexity of graph modeling and graph query languages:
    \begin{displayquote}
        We like the semantic explicitness of them [knowledge graphs], even if again, they're still hard to work with. -P9
    \end{displayquote}
    }
    

    

\end{itemize}

In addition the above affordances, participants told us they distinctly use KGs to perform data augmentation, generate concept maps or ontologies, and to organize libraries of data assets via data catalogs.

\subsection{Usage of Existing KG Tools}
\label{sec:tools}

    
    Our participants find success in several databases, tools, and methods for representing KGs:
    

\begin{itemize}[topsep=0pt, partopsep=0pt,itemsep=1pt,parsep=0pt]
      \item 7/19 use Neo4j with CYPHER;
      \item 6/19 use the Resource Description Framework with SPARQL;
      \item 4/19 use the NetworkX\cite{hagberg2008exploring} Python package;
      \item 2/19 use SQLite3 with SQL;
      \item 1/19 use spreadsheets and adjacency matrices.
\end{itemize}

    Our participants also use a combination of tools, libraries, and interfaces for visualizing their KGs: 
    \begin{itemize}[topsep=0pt, partopsep=0pt,itemsep=1pt,parsep=0pt]
      \item 7/19 use Neo4J Bloom\cite{neo4jbloom};
      \item 5/19 use Gephi\cite{bastian2009gephi};
      \item 4/19 use NetworkX;
      \item 4/19 use Cytoscape\cite{shannon2003cytoscape};
      \item 2/19 use D3.js\cite{d3js};
      \item 6/19 did not use any visualization tools.
\end{itemize}
    
    One participant explained the need to create custom visualization tools for Consumers: 

   \begin{displayquote}
        Typically I find interfaces like \textit{Gephi} good for your initial exploration, but once you start wanting to put together a dashboard for an end user to use\ldots you're going to find you want features in there that these tools don't have. I've always found that they're good to start with, and then I have to make custom tools after that. -P2
    \end{displayquote}
    
    We discuss the benefits and tradeoffs of these visualization tools for KGs in Section~\ref{sec:node-link-diagrams}, as well as in our supplemental material.

    


\section{Knowledge Graph Challenges}
\label{sec:findings-kg-challenges}

Our participants reported several broad challenges with using KGs in practice, many of which are fundamentally rooted in data sourcing and data quality issues. We outline the most common challenges experienced by our participants, and use these challenges to motivate directions for visualization research in Section~\ref{sec:findings-kg-opportunities}.

\subsection{Data Quality}
\label{sec:challenges-data}
The most common challenge faced by our participants (15/19) are problems surrounding data quality. These challenges include:

\begin{itemize}[topsep=1pt, partopsep=0pt,itemsep=1pt,parsep=1pt]
  \item Sparse or missing data\cite{toussaint2022troubles}, i.e. nodes or links that participants know should appear, but for one or more reasons do not.
  \item Incorrect or unverifiable data\cite{paulheim2017knowledge}, i.e. nodes or links that participants know should not be in the dataset.
  \item Obsolete data\cite{lissandrini2022knowledge}, i.e. nodes or links in the dataset are no longer valid or relevant.
  \item Duplicate entities\cite{gal2014uncertain}, i.e. multiple nodes or links in the dataset that actually should be combined into a single node or link.
\end{itemize}

The majority of these problems stem from incomplete or in-progress enterprise KGs. Data quality issues negatively impact AI/ML collaborations (e.g., data cascades\cite{sambasivan2021everyone}), making it difficult to account for a model's true robustness to missing data, noise, duplications, and so on. 

While open-source KGs may be ``complete'' and useful for testing AI/ML models, they can also be unrealistic when compared to a real-world KG: ``\textit{The drawback of using WikiData is that it’s \emph{too} good \ldots there are no holes you'd otherwise find in practice}'' (P1). 





\begin{itemize}[leftmargin=0pt,topsep=4pt, partopsep=0pt,itemsep=2pt,parsep=2pt]
    
    \item[]{\textbf{Manual Data Updates:} 
    As with many kinds of data sources, KGs incur problems related to manually entering, validating, and invalidating data. While some KGs can be automatically generated, many still require manual human data entries to curate\cite{wikidata2023stats}, 
    which can be extremely burdensome on Builders who create enterprise data catalogs:

    \begin{displayquote}
        When you have 10 thousand attributes, it’s not humanly possible to sit down and define all of them\ldots Then if a system changes, the tags become obsolete.
        -P18
    \end{displayquote}

    P19 described the challenge of validating KGs with millions of nodes, particularly when Consumers must manually perform this validation:
    
    \begin{displayquote}
        Domain scientists need to manually validate whether this extraction makes sense or not, whether it’s completely nonsense or it is correct. So we have to use human experts to validate randomly selected sample data. -P19
    \end{displayquote}

    P16 shared they must periodically
    rebuild their KG 
    from updated source data. 
    During this rebuilding process, P16 and their team know that certain nodes and connections from the source data are invalid, however, it is difficult to manually annotate and `integrate' these 
    invalidations:
    
    \begin{displayquote}
    We often see you know, an association that doesn't make sense, or that we've already invalidated internally\ldots
    it would be nice to have an easy way to basically flag that
    for all future versions of the KG that are built. -P16
    \end{displayquote}
    }

    \item[]{\textbf{KG Entity and Path Challenges:} Challenges strongly associated with KGs include entity disambiguation and chokepoints in path discovery. Several (9/19) participants said they face challenges with entity ambiguity, in which a node or edge has multiple meanings in the KG:
    
    \begin{displayquote}
        One problem is entity disambiguation. 
        We take the data as is, we're not sure that two nodes are actually different. 
        -P8
    \end{displayquote}
    
     From the participants using KGs for path discovery, 4/9 
     have problems with chokepoint nodes through which many paths converge then diverge. When running path-finding algorithms to discover new connections, densely connected nodes in between the target nodes can grossly inflate the number of discovered paths, leading to irrelevant outputs:
    
    \begin{displayquote}
        A big problem is some of these intermediate layers are a lot smaller than the layers they're connected to, so they're 
        chokepoints. If you do that two step linkage between the two nodes, you end up with probably a lot of things that are irrelevant and more nodes than you actually want\ldots
        it would be super cool if we could use machine learning to infer the appropriate linkages. 
        -P9
    \end{displayquote}
    
   Some participants told us their team is experimenting with hyperedge representations, i.e. edges that connect more than just two nodes, in an attempt to avoid path chokepoints and problematic path convergence. 
    }

\end{itemize}

\subsection{Querying}

Querying is the most challenging problem faced by both KG Analysts and Consumers (11/19), particularly because each KG representation method typically has its own unique querying language\cite{hogan2021knowledge}.
Learning a graph query language is also difficult for end users:


\begin{displayquote}
    It's a hard sell to get them [end users] to spend the time to invest in learning those query languages without knowing that they're going to get something out of it. 
    -P16
\end{displayquote}

Even though KG Buidlers and Analysts remarked that end users should not have to (nor want to) learn a KG query language, one KG Consumer told us that they still need the ability to ``ask the KG'' questions for their own downstream tasks:

\begin{displayquote}
    As an end user who doesn’t write SPARQL, I would like to ask [the KG] who my best customers are. I'd like to visually explore the graph to determine that, since I know there are many possible answers to that question. -P13
\end{displayquote}

\begin{itemize}[leftmargin=0pt,topsep=4pt, partopsep=0pt,itemsep=2pt,parsep=2pt]
    
    \item[]{\textbf{Lack of Interim Results:} 
    Often as a user is developing a query, they need an interim subset of results to determine if the query is pulling relevant information. However, many querying systems instead wait until all the information is ready before returning the full set of results:

    \begin{displayquote}
        It was very frustrating because you'd construct a query, it would take like 15 minutes to load, and then you'd get no results after it finished. So having some type of interim result\ldots like `kept alive' with examples of the stuff it's bringing back, that kind of interaction would be helpful. At least just to make sure that you're on the right track. -P4
    \end{displayquote}

    Long wait times caused by computation is frustrating for users\cite{shneiderman1984response} and interrupts their analysis workflows\cite{liu2014effects}. Work in progressive visualization\cite{angelini2018review} could alleviate similar issues for KGs. 
    }


    

\end{itemize}




\subsection{Socio-Technical Problems}
Two of our participants mentioned that many of their challenges have both social and technical aspects that stem from difficulties in interpersonal communication and collaboration. 

\begin{itemize}[leftmargin=0pt,topsep=4pt, partopsep=0pt,itemsep=2pt,parsep=2pt]
    
    \item[]{\textbf{Incomplete Understanding of End Users' Needs:} 
    P17 has observed that many people are drawn towards creating KGs before properly understanding the overarching use case and needs of the end users:
    
    \begin{displayquote}
        People always want to go build a knowledge graph of everything\ldots They've defined success from a technical point of view. But what are you going to do with it? Why are you doing it? Who's going to go use it? What's the value it's going to produce? -P17
    \end{displayquote}    
    
    P18 (a KG Consumer) told us a similar story about a developer (Builder) who unnecessarily over-complicated the construction of a KG:
    
    \begin{displayquote}
        I have a firm belief that a developer added all those features [to the KG] because they thought, `the more features, the better,' instead of considering what we actually needed for analysis. -P18
    \end{displayquote}
    
    In the end, an enormous amount of time and effort is spent to create a KG that might not necessarily have utility for its users. P18 was adamant that a simpler version of their company's KG would have better met the end users' needs -- leading to its wider adoption.
    }

\end{itemize}

\begin{itemize}[leftmargin=0pt,topsep=4pt, partopsep=0pt,itemsep=2pt,parsep=2pt]
    
    \item[]{\textbf{Non-standardized Nomenclature:} 
    Another challenge is a lack of standardized nomenclature, in which different groups of people may use one word to describe multiple meanings, or alternatively use various terms to describe the same concept:
    \begin{displayquote}
        There’s this concept of profit. What does profit mean? Well, it’s actually this complicated math that's not in the source. 
        So we go talk to different people, and they're gonna have different answers. 
        One word can mean multiple things to multiple people.
        -P17
    \end{displayquote}
    }
    
    
    \item[]{\textbf{Organizational Politics and Unsustainability:} 
    Across enterprise settings, KGs can fail to be adopted for political reasons, or fail to be maintained due to its long-term unsustainability:
    \begin{displayquote}
        Was our knowledge graph successful? No. The reason for failure was more political. You need funding and resources from leadership, but the interest died out. 
        -P18
    \end{displayquote}
    Organizational issues are often cited as a major reason for AI and ML models failing to be adopted in industry settings\cite{passi2018trust}. 
    }

    \item[]{\textbf{Too Many Hammers, Not Enough Talking:} While research continues to focus on optimizing knowledge graph databases and query languages, P17 told us that many technological problems are already solved. Instead, P17 believes computer scientists need to be open to addressing the social problems related to KGs:
    
    
    
    \begin{displayquote}
    We can build faster graph database systems. Is it intellectually challenging to go do that? Definitely. But is that going to
    push the barrier for the world to take unknowns and turn them into knowns? Honestly, no. 
    We don’t need more hammers. We need to go figure out how to use the hammers\ldots 
    This is where computer scientists can get uncomfortable, doing qualitative methods. 
    -P17
    \end{displayquote}
    }

\end{itemize}

\subsection{Current KG Visualization Designs}
\label{sec:node-link-diagrams}
A major point of discussion in our interviews was how current KG visualization designs either meet or fail to meet users' needs. 
By far, node-link diagrams\cite{herman2000graph} (NLDs) were the most commonly used KG visualization across all three KG user personas (18/19). However, we find that NLDs have shortcomings for the (albiet) many challenges they tackle. We discuss those challenges below.

\begin{itemize}[leftmargin=0pt,topsep=4pt, partopsep=0pt,itemsep=2pt,parsep=2pt]
    
    \item[]{\textbf{Lack of Scalability for Visual Sanity Checking:} Node-link diagrams (NLDs) are commonly used by KG Builders (9/12) as ``sanity checks" to 
    ``\textit{make sure nothing weird is going on\ldots that the graph is connected as expected}'' (P14).
    For creating NLDs, most of our participants used Gephi\cite{bastian2009gephi}.
    %
    %
    When asked about the limitations of node-link diagrams for sanity-checking, participants told us that scalability was the biggest issue: ``\textit{I don't want to have to see the full graph\ldots I'd just like to see something like, a quick sanity check}'' (P6). 
    
    Scalability is an issue at two levels. 
    First, when the knowledge graph is dense, it can be difficult to make sense of the resulting NLD. To alleviate this problem, an interactive NLD (e.g., force-directed layout\cite{gibson2013survey, suh2020phFDLs}) is used; however, when the KG is very large, it can be computationally difficult to render the entire graph -- an optimization problem that is still an ongoing area of graph visualization research\cite{von2011visual, gomez2018visualizing}. 
    }
    
    

    \item[]{\textbf{Lack of Efficacy for KG Consumers:} 
    %
    %
    Many (12/19) participants explicitly mentioned that NLDs are impossible to interpret by end users at the scale of thousands, millions, or billions of nodes.
    Several (9/19) participants criticized NLDs for quickly turning into ``hairballs,'' making it difficult for end users to digest meaningful information:
    
    
    \begin{displayquote}
        Because graphs are very visual, I hypothesize that there's something socially in our brain that automatically says, I want to see it. 
        I hear this all the time, `I just want to see it.' Okay, so you see this small graph,
        so what? `Show me something bigger.' Then it turns into a hairball. 
        This is the story of graph visualization. 
        -P17
    \end{displayquote}

    P12 told us about an interactive graph interface that his development team built, only for end users to reject it in favor of a table diagram:
    
    \begin{displayquote}
        We put a lot of software developers' effort into this GUI that showed our analysis as a graph to the user. But the results that came back always looked like a ball of yarn, unstructured\ldots Users couldn't make sense of it. 
        In the end, 
        they preferred a table. We played around with different ways to clean up and summarize the graph, but we never found a good way to visualize the ``graph-ness'' of the data in a way that the users could navigate. -P12
    \end{displayquote}

    From our interviews, we believe KG Consumers tend to prefer tables over other representations due to: (1) their simplicity and familiarity across multiple domains, and (2) the Consumer's task at hand is often straightforward (e.g., data retrieval). This is in contrast to Builders and Analysts, who tend to prefer more comprehensive visualizations (e.g., NLDs) when exploring the data or completing more complex tasks. 
    
    Regardless of the shortcomings of NLDs, two participants specifically mentioned that they make for good eye candy: ``\textit{I will argue that they make very pretty pictures\ldots they're great slide decoration}'' (P13).
    
    
    }

\end{itemize}

\section{Visualization Opportunities for Knowledge Graphs}
\label{sec:findings-kg-opportunities}
Finally, we present directions for visualization research that can begin to alleviate many of the challenges identified in Section~\ref{sec:findings-kg-challenges}. Where appropriate, we include participant quotes or references to related literature that motivates each recommendation.

\subsection{Graph-Abstracted Visualizations}
\label{sec:vis-opportunity-graph-abstracted-visualizations}

A recent Dagstuhl report\cite{klein2022bringing} discusses future directions of research at the intersection of graph databases and network visualization. With respect to the current limitations identified in this work, it is critical that a relevant design space for knowledge graph visualizations (and their end users) is explored and contributed. Building on previous graph visualization work can provide a starting point\cite{task_taxonomy, 2016_eurovis_pathfinder, 2018_infovis_juniper}.

Beyond standard network visualizations, some (5/19) participants advocated for KG applications and visualizations that are \textit{not} graphs:

\begin{displayquote}
    Just because you have a knowledge graph doesn't mean the visualization has to be a graph\ldots Under the hood, the end user doesn't even have to know it's there. But they're getting the benefits of having it there in terms of improved results\ldots
    I've 
    come to see that the most effective way for non-technical experts to engage with KGs is through specific applications that are powered by a KG, rather than directly tangling with a ball of yarn visualization 
    -P13
\end{displayquote}

P16 found that end users are not only confused by NLDs, but trust the analysis results less when they are presented as a graph visualization:

\begin{displayquote}
    In our experience, the more we can shelter the end user from the underlying graph structure, the better their willingness to interact with the data and accept the results that come out of it. As soon as the level of complexity of the graph reaches a certain level on the screen, users really tend to shut down and not trust any of it. -P16
\end{displayquote}

P15 agreed, telling us that, ``\textit{there’s a big difference between the format that machines want to read data, versus the format that humans want to read data.}'' While KGs store complex data (knowledge) that is interpretable to both humans and machines, this feedback suggests that users do not necessarily prefer to \textit{see} this knowledge in graph form. 


\medskip 

\noindent 
\textbf{What if I have to (or want to) use a graph visualization?} As discussed in Section~\ref{sec:node-link-diagrams}, many of our participants still have to use NLDs (or general graph visualizations) for a variety of use cases and applications, particularly Builders and Analysts. Our participants identified several capabilities, often lacking in current tools, that should be supported when interacting with or consuming KG visualizations: 

\begin{itemize}[topsep=4pt, partopsep=0pt,itemsep=2pt,parsep=2pt]
    \item The ability to immediately begin analysis at the user's desired point or region of interest (i.e. drilled drill down into the KG).
    \item The ability to filter, bundle, condense, collapse, or expand areas (regions) of the KG during open-ended exploration.
    \item The ability to switch views, while maintaining context, depending on the KG data type (e.g., from graph view to table view).
\end{itemize}

\subsection{Balancing Digestibility and Discoverability}
\label{sec:vis-opportunity-digestibility-discoverability}

One of the advantages to node-link diagrams is the ability for users to traverse the KG from node to node to discover new information. However, users find NLDs ineffective in practice because the scales of KGs 
they work with (Section~\ref{sec:users-use-cases}).
Consequently, there is a need for visual interfaces that balance both digestibility and discoverability, to allow users to properly process information and explore the KG:

\begin{displayquote}
    I think the key with [KG interfaces] is not to lose the pivotability and discoverability that is fairly unique to graphs\ldots If you can maintain this sort of continuous reference to the graph that allows for that organic discovery process as opposed to like static results\ldots that could be really good.
    -P13
\end{displayquote}

\begin{itemize}[leftmargin=0pt,topsep=4pt, partopsep=0pt,itemsep=2pt,parsep=2pt]
    
    \item[]{\textbf{Wikipedia as an EDA Tool:} 
    Four of our participants expressed that Wikipedia functions as an effective graph exploration tool, since Wikipedia simultaneously presents detailed information about a specific article (i.e.\ node) and also embeds hyperlinks to other related articles (i.e. edges connecting to another node). In this interface setup, users can both explore the graph structure of Wikipedia while also gaining valuable insights of their choosing by interacting with articles. We discuss the design of potential KG interfaces to support seamless knowledge discovery in Section~\ref{sec:vis-opportunity-interfaces}.}

\end{itemize}

 \begin{figure} 
    \centering
     \includegraphics[clip, trim=5pt 6.2cm 14.2cm 2pt, width=.48\textwidth]{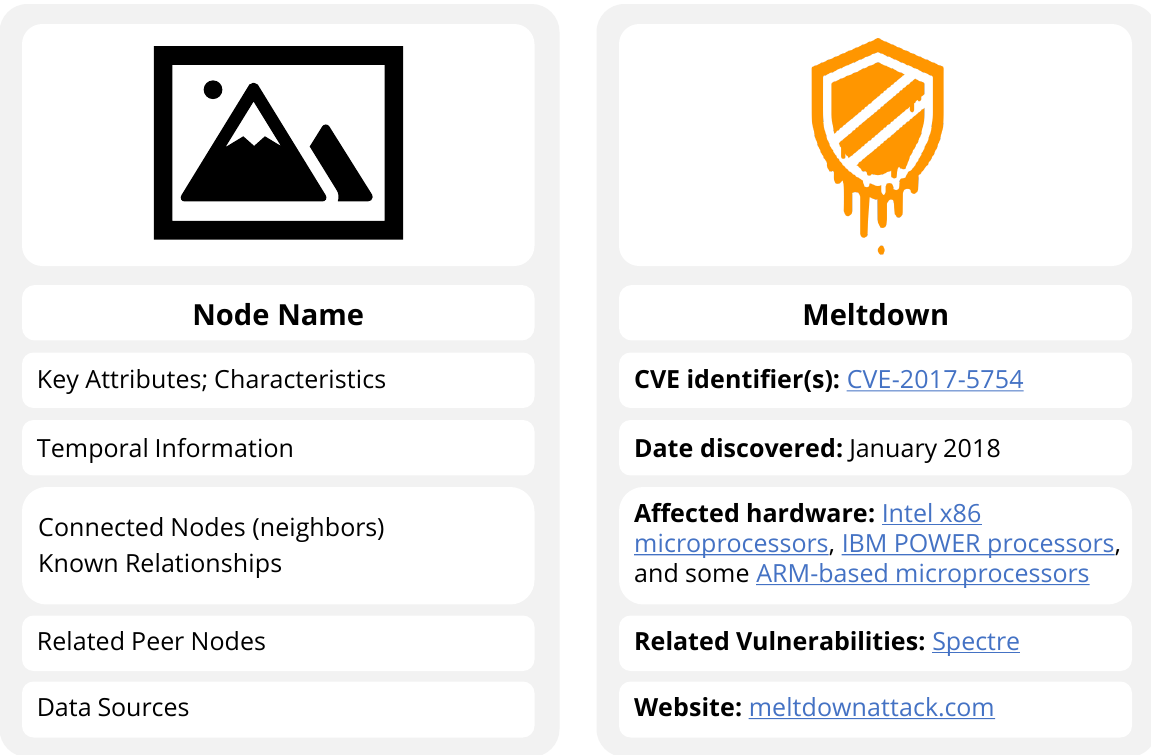} 
    \caption{%
    Left: an example knowledge card template with node and edge information that may be relevant to a KG end user.
    Right: an example knowledge card of a cybersecurity vulnerability that we iterated on with one of our participants to understand what might be useful to a cyber analyst. We describe knowledge cards in Section~\ref{sec:findings-knowledge-cards}.
    }
    \label{fig:knowledge-cards}
\end{figure}

\begin{itemize}[leftmargin=0pt,topsep=4pt, partopsep=0pt,itemsep=2pt,parsep=2pt]
    \label{sec:findings-knowledge-cards}
    \item[]{\textbf{Contextual Knowledge Cards:} 
    Similar to a \textit{baseball card}, a knowledge card can give a high-level summary of the most essential data for that particular node or entity in the KG. 
    Five of our participants said they use similar visualizations to deliver KG context to Consumers:

    \begin{displayquote}
        A knowledge card can be a really powerful as a visualization in use cases that I want to understand the context of something. This is where knowledge graphs are really powerful, where they have the advantage over like traditional relational databases\ldots being able to understand context and relationships. 
        -P13
    \end{displayquote}

    A template for the basic information expected in a knowledge card is provided in Figure~\ref{fig:knowledge-cards}. To create this template, we had follow-up calls with P5.
    During these calls, we iterated on knowledge cards that would be useful for their particular use case. In general, the information needed was highly specific to the domain and use case. General information included: an image of the entity being queried from the KG, key attributes or identifying characteristics of the entity (including temporal data e.g., updates to the entity or when the entity was created), known paths, relevant entities, and data sources or credentials. 
    
    
    }
    
\end{itemize}

Connecting this research opportunity to Section~\ref{sec:vis-opportunity-graph-abstracted-visualizations}'s, another interesting direction could include the investigation of using knowledge cards as the ``nodes'' in a node-link diagram. Further, when clustering groups of nodes in the KG for a `global view,' a knowledge card could represent a high level abstraction for collections of nodes (e.g., a card representing a cluster of universities or sports teams).

\subsection{KG-Based Interfaces that Support Organic Discovery}
\label{sec:vis-opportunity-interfaces}

There are a variety of opportunities for visual interfaces to align with the capabilities of KGs, thereby facilitating data discovery and exploration.
In particular, interfaces that allow end users to ``ask the KG questions'' that they would not have thought to ask in the first place:

\begin{displayquote}
When I’m looking at an entity in the knowledge graph, I want all the interesting questions and answers about that entity available\ldots So, for example, if there’s a drug undergoing a particular clinical trial, I want to be able to quickly have all the interesting properties about that drug on a page\ldots [to support] very fast discovery of insights about individual instances of data that I probably wouldn’t have found without the tool showing me.
-P15
\end{displayquote}

We identified two types of exploration use cases for KG visual interfaces: those for unguided, unstructured, and organic exploration (i.e. ``\textbf{open-ended KG exploration}''), and those for directed and targeted exploration (i.e. ``\textbf{goal-oriented KG exploration}''):

\begin{displayquote}
One is a very exploratory type of search, just navigating, clicking around and trying to learn things with no clear goal. Have you ever clicked on something on Wikipedia, 
then you end up going down this rabbit hole? Did you have an objective? No, but you did a bunch of work, and you probably learned something. 
The other [type] is something very specific\ldots  
Think of `Googling' something. This is a search problem, but with intention. -P15
\end{displayquote}

We believe there are two directions for KG-based visualization interface design to support both types of exploration. 

\begin{itemize}[leftmargin=0pt,topsep=4pt, partopsep=0pt,itemsep=2pt,parsep=2pt]
    
    \item[]{\textbf{Visual Interfaces Built on Top of KGs:} One direction is for interfaces built ``on top of the KG,'' that is, users interact with an interface to explore and ask the KG questions directly. In these interface designs, interactive visualization can help users make sense of the KG and query it, possibly without any knowledge of the KG's inherent query language (e.g., similar to using Wikipedia or Google's search engine). There has been prior work done in the KG community to support KG exploration through query-building GUIs\cite{grafkin2016sparql, ferre2017sparklis}. Similar efforts have been made to lower the barrier for querying relational databases, such as NoSQL\cite{han2011survey}, as well as natural language queries (e.g., NL-to-SQL)\cite{Gan2021NaturalSM}, and natural language interfaces (NLIs)\cite{Aurisano2016Articulate2, Narechania2020NL4DVAT}. While NLIs are becoming more popular in the visualization research community (e.g.,\cite{mitra2022facilitating, huang2023flownl}), we see an untapped opportunity to integrate these techniques with KG exploratory visualization tools.}

    \item[]{\textbf{Using KGs to Augment Exploratory Visualization Tools:} Another direction is utilizing KGs altogether to enhance current visualization interfaces. With the growing popularity of public KGs\cite{waagmeester2020wikidata}, visualization system designers can consider adding the ability for users to ``connect'' to a KG relevant to their own domain or task. For example, given a generalizable exploratory visualization tool like Tableau or Voyager\cite{2017-voyager2}, an additional widget to ``ask a KG'' may be implemented to enrich exploration with data that the user is not currently connected to, increasing in-situ discovery and data integration (similar to\cite{Cashman:2020:CAVA}). Given a KG's semantic richness and well-defined (often hierarchical) structure, KGs could seamlessly provide additional context, annotations, supplementary visualizations, etc. to an analysis session.}
\end{itemize}

\subsection{KGs for Explainability}
The semantic nature of KGs (Section~\ref{sec:benefits-of-using-kgs}) can be leveraged for explainable AI (XAI) to help model creators debug their model during training, and also to help end users trust a model's predictions\cite{tiddi2020knowledge}. This is particularly valuable for deep neural networks, which are often regarded as black boxes with limited interpretability\cite{Rudin2019Why}. Lecue\cite{lecue2020role} describes opportunities for using KGs to help encode the semantics of inputs, outputs, and their properties in a neural network.

\begin{itemize}[leftmargin=0pt,topsep=4pt, partopsep=0pt,itemsep=2pt,parsep=2pt] 
    
    \item[]{\textbf{XAI for Model Debugging:} Three of our participants wished there were tools to help them debug and improve ML and KG model training:}
\end{itemize}

\begin{displayquote}
    Since machine learning is mostly my focus, visualization would have been helpful\ldots When the data is in this intermediate knowledge graph form, it's really hard to debug and visualize performance\ldots it's hard to tell like, is my model doing worse because the data is different, or because the data is exactly the same except for some small detail? -P9
\end{displayquote}



\begin{itemize}[leftmargin=0pt,topsep=4pt, partopsep=0pt,itemsep=2pt,parsep=2pt] 
    
    \item[]{\textbf{XAI for KG Analysts and Consumers:} 
    KGs can serve as an effective tool for providing predictive explainability for Analysts and Consumers. For instance, in image analysis, saliency maps can highlight the specific areas in the input image that a neural network focused on, providing insights into why the model arrived at a particular prediction\cite{simonyan2014deep}. Similarly, ML models trained on KGs could offer meaningful semantic explanations for their predictions by highlighting the relevant nodes and edges that played a role in shaping their predictions. By leveraging KGs, we can empower users to understand the reasoning behind a model's predictions and build greater trust in its outputs.}

\end{itemize}

\noindent 
As an illustrative example, P9 is developing an anomaly detection algorithm that evaluates the semantic proximity of objects using a KG connecting household objects and their locations. Their model should ideally be able to detect that a hammer is an anomaly in the context of a kitchen. In their case, the KG helps ``explain'' the detected anomaly: 

\begin{displayquote}
    Hammers are normally in a shed. And our hope was basically that by using a source of context, we could do anomaly prediction for things like, should a hammer be in the kitchen? The answer is no here. So the way we chose to get that context was through a knowledge graph. -P9
\end{displayquote}

P9 described two types of explanations that the KG can provide for why the model classifies a hammer in a kitchen as an anomaly: 

\begin{enumerate}[topsep=2pt, partopsep=0pt,itemsep=1pt,parsep=2pt]
    \item Nodes in the KG like pots and plates appear in this context, but are not closely related to hammers in the KG.
    \item Nodes in the KG like wrenches and shovels are closely related to hammers, but do not appear in this context.
\end{enumerate}

We posit that these kinds of contextual explanations can be important for KG-based XAI, and potentially powerful for visual analytics.

\subsection{KG Timelines: Tracking Evolutions Over Time}
\label{sec:kg-evolution}
We identify two distinct temporal-based directions for knowledge graph visualization research.

\begin{displayquote}
    Another peculiarity of my data is the entities are all time stamped roughly. So it would be interesting to see the evolution of products or entities over time. - P9
\end{displayquote}

\begin{itemize}[leftmargin=0pt,topsep=4pt, partopsep=0pt,itemsep=2pt,parsep=2pt]
    
    \item[]{\textbf{Visualizing Multi-Attributed Time-Series KG Data:} 
    First, there is a need to consume and understand time series, temporal, or ``timestamped'' data from knowledge graphs, e.g., the data in EventKG\cite{gottschalk2018eventkg}. 
    We observe the need for new visual designs and interfaces that allow users to precisely (and organically) navigate and consume temporal data, events, and relations in a large-scale KG. Work similar to Brehmer et al.'s\cite{brehmer2016timelines} could be done to contribute design spaces for multi-attributed temporal (time-series) data, since the nodes and edges of knowledge graphs typically contain multiple attributes.} 
    
    \item[]{\textbf{Tracking KG Data Evolution and Authoring Changes:} KG users need visualization solutions that track how the knowledge graph has changed over time, similar to previous visualization work in tracking software changes\cite{storey2005use, yoon2013visualization}. Specifically, users need help tracking \textit{and} validating in what capacity the KG has been assigned additional information, e.g., through new edges (relations) added. 
    Users also need to know whether new information drastically changes their mental model of the KG, or makes their analyses out of date. For example, if an Analyst is curating a report using news articles contained within a KG, it is important for them know whether their current state of information is obsolete. 
    As discussed in Section~\ref{sec:challenges-data}, some users also do not have an effective method of annotating data invalidations to omit certain nodes or links when they rebuild KGs from source data.}
\end{itemize}

\subsection{Mapping Dynamic Data onto Static Views}

One of the challenges associated with using NLDs for large KGs is that the algorithms used to generate ``nice'' and computationally fast layouts (e.g., FDLs\cite{gibson2013survey}) often calculate the node and link positions dynamically (or stochastically). Consequently, the graph layout can change drastically each time the KG visualization loads, which can confuse users who must reorient themselves after each change.

One possible solution was posed by P12, whose team of biologists found success in always visualizing KG data on top of the Roche Biochemical Pathways\footnote{\url{http://biochemical-pathways.com/\#/map/1}} diagram, a standardized graph visualization detailing various biochemical processes:


\begin{displayquote}
    There are networks in biology that people [biologists] are already familiar with that are best visualized as a graph. 
    The graph becomes static, then you can load data and project it onto that map. This lets me highlight parts in the network
    that were active in an assay, or played a role in diabetes.
    -P12
\end{displayquote}

For example, the same method is used when visualizing navigational directions. The network of roadways remains static, while routes and icons can be overlaid on top. While this KG visualization opportunity may be use case specific, it could support Consumers who require the context of their own domain to extract insights from the KG.

\subsection{KG Schema Creator and Enforcer}
\label{sec:vis-opportunity-schema}

A good, consistent knowledge graph hinges on a good, consistent knowledge graph schema\cite{abu2021domain}. Interview participants (typically KG Builders) told us that creating a reliable schema can take several months at a time, which halts the actual development process of the KG database:

\begin{displayquote}
    I think another thing that's very much missing from the [KG] landscape is how to even build the graph to begin with, 
    how to put together your schema\ldots 
    It can make querying impossible if you don't build it correctly. Like if I build it in \textit{this} way, I can get \textit{this} information out of it. But if I build it in this other way, I won't ever be able to do \textit{this} query.
    -P2
\end{displayquote}

\begin{itemize}[leftmargin=0pt,topsep=4pt, partopsep=0pt,itemsep=2pt,parsep=2pt]
    
    \item[]{\textbf{KG Schemas as Visual Maps:} 
    There are many graph and tree visualizations that could act as a starting point for schema visuals\cite{herman2000graph}. KG Builders need concise but detailed views of the schemas they are creating, maintaining, and iterating on for their knowledge graph. Moreover, visual designs should consider that schemas may change over time: attributes can be added to nodes or edges, new relations can be created, while others may be removed entirely. A good way to highlight how a schema has changed over time should be integrated.} 
    
    \item[]{\textbf{Interactive Schema Builder \& Enforcer:} Interview participants that spend months building a schema frequently use tools like \textit{Visio}\cite{helmers2015microsoft} to create the schema framework or template (i.e. a schema visual map). The common usage of Visio should make clear what is needed for an interactive schema builder: flexibility, customizability, and a multitude of design tools. However, what is lacking in a generic tool like Visio is the ability to: (1) preview the schema, (2) integrate directly into a KG workflow, (3) enforce types and constraints in the schema (e.g.,\cite{bonifati2022pg}). In an ideal tool, the KG Builder could also preview how different queries could be accomplished given the current schema design. }

\end{itemize}


\section{Discussion}
\label{sec:discussion}

\subsection{Domain-Specific KG Visualization Designs}
\label{sec:discussion-domain-specific}
We began these interviews with the assumption that there is a common challenge faced by most KG practitioners which could be addressed by a generalizable KG visualization system. Instead, we quickly found that -- even though our participants did share common challenges -- their domain-specific needs could not be met by a single visualization solution. Many (14/19) participants explicitly stated they believe that \textbf{visualization challenges for KGs are domain specific for end users}.




\begin{displayquote}
    It's probably going to be really hard to find visual metaphors that work for everybody\ldots and for different tasks, right? 
    I think the domain
    will prioritize how you visually present the data, and then that's what a useful visual metaphor for the data will be.
    -P10
\end{displayquote}


As P2 told us: ``\textit{It's going to be hard to make a generalization unless you know the exact use cases}.'' Creating effective KG visualizations will require additional formal or informal user studies\cite{selmdmair2012dsm} to understand the end users' data domain, applications, questions, and needs.

\subsection{What Defines a Knowledge Graph?}
\label{sec:discussion-kg-definition}
There is often confusion for what makes a data model a KG beyond storing nodes and edges, a commonality across any graph data representation. 
We asked our interview participants to provide their own definition and criteria for a knowledge graph, the full list is included as supplemental material. 
Based on our own participants' collective definitions of a KG, we offer the following description:
\begin{displayquote}
    \textit{A data model representing entities as nodes, the multi-relationships between those nodes as edges, and properties defining them, such that humans \emph{and} machines can easily understand the nuances of that data due to its semantics.}
\end{displayquote}
The most important criteria of a KG identified by our participants were: (1) its ability to store different types of nodes (or entities), different types of semantic relationships (or edges) between those nodes, and the attributes (or properties) on them; (2) its ability to help both humans and machines understand what the data ``\textit{actually means}'' -- e.g., in the greater context of the data domain or use case. 
  
\subsection{Limitations \& Future Work}
\label{sec:discussion-limitations}

Additional work beyond our interview study is needed to further understand the role of visualization for knowledge graphs used in practice.  While KGs have been an active area of research in other communities (e.g., database systems\cite{lissandrini2022knowledge} and NLP\cite{chen2020review}), they have only recently become a target of study in visualization research (e.g.,\cite{Cashman:2020:CAVA, li2021kg4vis}). There are a multitude of opportunities for visualization research to leverage the semantic-richness of KGs, as well as application-driven research to augment current KG tools. We presented our participants with three tools\cite{latif2021visually, gottschalk2018eventkg, neo4jbloom} discussed in Section~\ref{sec:related}, with mixed feedback on their perceived helpfulness (see supplemental). Future work should investigate to what extent existing graph visualization research can be applied to KGs, as well as how accessible these systems are to practitioners.

For our study, we interviewed 19 practitioners across eight organizations, with roughly half the participants coming from the same FFRDC. We envision a wide array of future studies conducted to better understand the challenges and needs of KG practitioners -- similar to the ongoing user-centered research being done for AI/ML collaborations\cite{sambasivan2021everyone}. As many of our participants discussed with us, robust technical solutions have already been posed related to building and completing KGs\cite{rossi2021knowledge}, however, social challenges related the usability of KGs remains a large-scale issue in many collaborative settings. 

Another limitation of our study is the lack of KG Consumers interviewed: 17/19 participants were either KG Builders or Analysts. 
This is in part because Consumers tend to work with applications that are served from a KG ``under the hood,'' and might have been self-selected out of our interview solicitation process.
Consequently, many of the identified challenges and posed solutions in this study relate to practitioners' feedback who work more directly with a knowledge graph.
Future work should therefore be done to target the ultimate end users of KGs -- even those that might be unaware that they in fact use KGs.
Suresh et al. conducted similar research for ML stakeholders\cite{suresh2021beyond}.   
  
In general, we were not able to identify any single solution for ``the best'' visual encodings for KGs -- instead, we identified shortcomings of current designs. However, we believe these findings underpin the need for further research in visualization for KGs, and KGs for visualization. While we posed a variety of visualization research directions and possible designs in Section~\ref{sec:findings-kg-opportunities}, future work will need to address the validity of those suggestions. This also opens up visualization research in curating KG task taxonomies (similarly, to compare and contrast to\cite{task_taxonomy}), design guidelines, and potential KG design spaces. 

\section{Conclusion}
\label{sec:conclusion}

We presented an interview study with 19 practitioners in industry and academic settings across eight organizations who regularly use knowledge graphs. From our interviews, we distilled common knowledge graph practices, uses cases, and tools frequently used by practitioners. We identified three personas for the users of KGs: (1) \textit{Builders} who create and maintain KGs, (2) \textit{Analysts} who explore and analyze the data in KGs, and (3) \textit{Consumers} who use the insights from KGs for downstream tasks. From those personas, we discussed how each KG user has distinct expertise, tasks, and visualization needs. Overall, we found a gap in current tools and visualization methods (e.g., the usage of node-link diagrams for representing and visually communicating large KGs) for the challenges experienced by interview participants. Based on these collective findings, we outlined several directions for visualization research to enable better KG maintenance, ``open-ended'' and ``goal-oriented'' data discovery, analyses, and collaborations. 


\acknowledgments{
We sincerely thank each of the practitioners who took the time to participate in our interview study, and the reviewers for their insightful feedback on improving the quality of our paper. This work was supported by National Science Foundation grants IIS1452977, OAC-1940175, OAC-1939945, OAC-2118201, NRT-2021874.

\medbreak
\noindent 
DISTRIBUTION STATEMENT A. Approved for public release. Distribution is unlimited. This material is based upon work supported by the Department of the Air Force under Air Force Contract No. FA8702-15-D-0001. Any opinions, findings, conclusions or recommendations expressed in this material are those of the author(s) and do not necessarily reflect the views of the Department of the Air Force.
}
\bibliographystyle{abbrv-doi}

\bibliography{main}

\end{document}